\documentclass[pre,aps,twocolumn,nofootinbib]{revtex4-1}

\usepackage{graphicx}
\usepackage{amsmath}
\usepackage{latexsym}

\newcommand{\bigsum}[1]{{\displaystyle \sum_{#1}}}
\newcommand{\dispderiv}[2]{{\displaystyle
                            \frac{\partial #1}{\partial #2}}}
\newsavebox{\tallguy}
\savebox{\tallguy}{\mbox{\rule{0ex}{2.25ex}}}
\newcommand{\tr}{\mbox{Tr} \, }
\newcommand{\ket}[1]{ \usebox{\tallguy} \left | #1 \right \rangle}
\newcommand{\bra}[1]{ \left \langle #1 \right | \usebox{\tallguy}}
\newcommand{\amp}[2]{
    \left \langle #1 \left | #2 \right. \right \rangle}
\newcommand{\proj}[1]{\ket{#1} \! \bra{#1}}
\newcommand{\outerprod}[2]{\ket{#1} \! \bra{#2}}
\newcommand{\ave}[1]{\left \langle #1 \right \rangle}
\newcommand{\hilbert}{\mathcal{H}}
\newcommand{\absolute}[1]{\left | #1 \right |}
\newcommand{\conj}[1]{#1^{\ast}}
\newcommand{\oper}[1]{\bm{#1}}
\newcommand{\doper}[1]{\bm{#1}}
\newcommand{\oforder}[1]{\mathcal{O} \left ( #1 \right )}
\newcommand{\onehalf}{\mbox{$\frac{1}{2}$}}

\begin{document}

\title{Probability current and thermodynamics of open quantum systems}
\date{\today}

\author{Benjamin Schumacher}
\affiliation{Department of Physics, Kenyon College, Gambier, OH 43022}
\email[Corresponding author: ]{schumacherb@kenyon.edu}
\author{Michael D. Westmoreland}
\affiliation{Department of Mathematics and Computer Science, Denison University, Granville, OH 43023}
\author{Alexander New}
\affiliation{Department of Mathematical Sciences, Rensselaer Polytechnic Institute, Troy, NY 12180}
\author{Haifeng Qiao}
\affiliation{Department of Physics and Astronomy, University of Rochester, Rochester, NY 14627}

\begin{abstract}
	This paper explores the generalization of the concept of a ``probability current'',
	familiar from wave-function quantum mechanics, to quantum
	systems with finite-dimensional Hilbert spaces.  The 
	generalized definition applies both to isolated systems evolving 
	via the Schr\"{o}dinger equation and to more general open
	systems obeying the Lindblad master equation.  We establish
	several properties of the probability current and explore its
	relation to thermodynamic heat and work.
\end{abstract}

\maketitle

\section{Introduction}

Consider a quantum system described by a Hilbert space $\hilbert$
of finite dimension $d$.  A non-degenerate observable is described by
an orthonormal basis $\{ \ket{n} \}$ for $\hilbert$, each basis vector
associated with a measurement outcome.  If the system
state vector is $\ket{\psi}$ then the probability of the $n$th basis
state is 
\begin{equation}
	P_{n} = \absolute{\amp{n}{\psi}}^2 \label{eq-bornrule} .
\end{equation}
As the system state evolves, this probability changes, but the total
probability remains 1.  That is, probability is ``conserved''.

We can express this idea by defining a {\em probability current}
$J_{nm}$, which is interpreted as the net rate at which probability
``flows'' from state $m$ to state $n$.  The probability current $J_{nm}$
is real and antisymmetric---that is, $J_{mn} = - J_{nm}$.  (As a corollary,
$J_{nn} = 0$ for any basis state $n$.)  
The net rate of change of $P_{n}$ will be given by
\begin{equation}
	\frac{dP_{n}}{dt} = \dot{P}_n = \sum_{m} J_{nm} .
	\label{eq-probabilityflow}
\end{equation}
Such probability flows are familiar from classical master equations \cite{Gardiner1985}.
There, $P_{n}$ represents the classical probability that a system
is in state $n$, and $W_{nm}$ represents the conditional rate at
which the system jumps from state $m$ to state $n$.  Then
\begin{equation}
	\label{eq-classicalmastereqn}
	\dot{P_{n}} = \sum_{m} \left (  W_{nm}P_{m} - W_{mn}P_{n} \right ) .
\end{equation}
For this sort of system, the classical probability current
$J_{nm} = W_{nm} P_{m} - W_{mn}P_{n}$ is real and antisymmetric,
and Equation~\ref{eq-classicalmastereqn} is exactly Equation~\ref{eq-probabilityflow}.
 
 Quantum probabilities do not change according to a classical
 master equation, but rather are governed by the underlying
 dynamics of the quantum state.  For an informationally isolated system
 \cite{SchumacherWestmoreland2010},
 the state vector $\ket{\psi}$ evolves according to the Schr\"{o}dinger
 equation
 \begin{equation}
	\label{eq-schrodingerpsi}
	\oper{H} \ket{\psi} = i \frac{d}{dt} \ket{\psi} ,
\end{equation}
where $\oper{H}$ is the system's Hamiltonian operator.
(Here, and in all subsequent expressions, we have set $\hbar = 1$.)
To make later generalizations more natural, we re-express 
Equation~\ref{eq-schrodingerpsi} in terms of the density operator
$\doper{\rho}$, which for a pure state is the projection $\proj{\psi}$
onto the state vector:
\begin{equation}
	\label{eq-schrodingerrho}
	\dot{\doper{\rho}} = \frac{d \doper{\rho}}{dt} = \frac{1}{i} \left [ \oper{H}, \doper{\rho} \right ] .
\end{equation}
This equation of motion also applies to mixed states of the system, for
which $\doper{\rho}$ is a positive trace-1 operator rather than a projection.
In any case, the probability $P_{n} = \bra{n} \doper{\rho} \ket{n} = \rho_{nn}$.

Silva \cite{Silva1992} derived a probability current $J_{nm}$ between basis states
for a system evolving according to Equation~\ref{eq-schrodingerrho}, applying 
his result to a one-dimensional model of electron transport.  More recently,
Roden and Whaley \cite{RodenWhaley2016} have presented a generalization that
applies to open-system dynamics, in which the density operator obeys a
master equation of a particular Lindblad type.  They use this to analyze how energy
flows between complexes of states, with particular attention to energy transport
in molecular processes such as photosynthesis.

The main purpose of this paper is to define a general probability current 
$J_{nm}$ for an evolving quantum system.  For Schr\"{o}dinger evolution,
we will show the close connection between our definition and the familiar 
probability current density $\vec{J}$ of wave-function quantum mechanics.  
We also show how to extend our definition to arbitrary Lindblad evolution,
and we note several important characteristics of the current $J_{nm}$,
including its invariance under the choice of Lindblad representation.
We then use the probability current to characterize the rates of
thermodynamic work and heat for a quantum system exchanging
energy with its environment.

\section{Schr\"{o}dinger evolution}

Suppose the state $\doper{\rho}$ of a quantum system evolves according 
to Equation~\ref{eq-schrodingerrho}, and let $\{ \ket{n} \}$ be a fixed 
orthonormal basis for $\hilbert$.  To define a suitable probability current $J_{nm}$,
we begin with the time rate of change of $P_{n} = \bra{n} \doper{\rho} \ket{n}$:
\begin{eqnarray}
	\dot{P}_{n} & = & \bra{n} \dot{\doper{\rho}} \ket{n} \nonumber \\
	& = & \frac{1}{i} \bra{n} \left ( \oper{H} \doper{\rho} - \doper{\rho} \oper{H} \right ) \ket{n} \nonumber \\
	& = & \frac{1}{i} \sum_{m} \Big( \bra{n} \oper{H} \proj{m} \doper{\rho} \ket{n}
		\nonumber \\ & & \quad \quad
		- \bra{n} \doper{\rho} \proj{m} \oper{H} \ket{n} \Big)
	\label{eq-pndothamiltonian}
\end{eqnarray}
This motivates us to define
\begin{equation}
	J_{nm} = \frac{1}{i} \left ( H_{nm} \rho_{mn} - \rho_{nm} H_{mn} \right ) .
	\label{eq-jdefhamiltonian}
\end{equation}
Since $\oper{H}$ and $\doper{\rho}$ are both Hermitian, $\conj{H_{nm}} = H_{mn}$ and
$\conj{\rho_{nm}} = \rho_{mn}$.  Thus, $J_{nm}$ is both real and antisymmetric, 
and it automatically satisfies Equation~\ref{eq-probabilityflow}.

Under what conditions can $J_{nm}$ be nonzero?  Equation~\ref{eq-jdefhamiltonian} gives
us important necessary conditions.  First of all, the Hamiltonian operator must have a 
nonzero matrix element $H_{mn}$.  This is a restatement of the well-known fact that 
transitions between basis states are produced by the off-diagonal elements 
of $\oper{H}$.

Less familiar is the corresponding condition on $\doper{\rho}$:  
$J_{nm} \neq 0$ only if the coherence $\rho_{nm} \neq 0$.
This means that, for example, in an incoherent mixture of 
$\ket{n}$ and $\ket{m}$ states, the probability current $J_{nm} = 0$.  Furthermore,
the postivity of $\doper{\rho}$ implies that
\begin{equation}
	\absolute{\rho_{nm}}^{2} \leq \rho_{nn} \rho_{mm} = P_{n} P_{m} .
\end{equation}
Thus, $J_{nm} \neq 0$ only if both $P_{n} > 0$ and $P_{m} > 0$.
This is the basis for the ``quantum Zeno effect'' \cite{SudarshanMisra1977}.
If $P_{n} = 0$ at any moment, then it is also true that
\begin{equation}
	\dot{P}_{n} = \sum_{m} J_{nm} = 0 .
\end{equation}
For Schr\"{o}dinger evolution, therefore, the departure of $P_{n}$ from 0 cannot
be linear in time.  Repeated measurements of the $\{ \ket{n} \}$ basis 
constantly reset $P_{n} = 0$ and thus suppress transitions to this state.

The idea of a ``probability flow'' is a familiar one from the quantum mechanics
of particles moving in one or more continuous spatial dimensions.  
Consider a particle of mass $\mu$ moving in one dimension.  The wave function
$\Psi$ yields a probability density $\mathcal{P} = \absolute{\Psi}^2$ for
the outcome of a hypothetical position measurement.  We define a 
{\em probability current density} by
\begin{equation}
    J = \frac{1}{2\mu i} \left ( \conj{\Psi} \dispderiv{\Psi}{x}
        - \Psi \dispderiv{\conj{\Psi}}{x}  \right ) .
    \label{eq-currentdensity1d}
\end{equation}
If the particle moves subject to a potential $U(x)$, its wave
function evolves according to the Schr\"{o}dinger equation
\begin{equation}
    - \frac{1}{2\mu} \dispderiv{^2 \Psi}{x^2} + U(x) \Psi  = i  \dispderiv{\Psi}{t} .
    \label{eq-schrodinger1d}
\end{equation}
Then the probability density satisfies a continuity equation
\begin{equation}
    \dispderiv{\mathcal{P}}{t} = - \dispderiv{J}{x} .
    \label{eq-continuity}
\end{equation}
These results are easily generalized to 3-D, in which the probability
current density is a vector $\vec{J}$.

How is our probability current $J_{nm}$ related to these standard
definitions?  We replace the continuous spatial coordinate $x$ with
a discrete lattice of points $x_{n}$, with adjacent points separated
by $\Delta x$.  In the state $\ket{n}$, the
particle is localized to the $n$th lattice point.  These localized states
form an orthonormal basis for the Hilbert space, so we can write any state as
\begin{equation}
	\ket{\psi} = \sum_{n} \psi_{n} \ket{n} ,
\end{equation}
where $\psi_{n} = \amp{n}{\psi}$.
We define the wave function $\Psi(x_{n}) = \psi_{n} / \sqrt{\Delta x}$, so that
\begin{equation}
	\sum_{n} \absolute{\psi_{n}}^2 = \sum_{n} \absolute{\Psi(x_{n})}^2 \Delta x
	\longrightarrow \int \absolute{\Psi(x)}^2 \, dx
\end{equation}
in the continuous ($\Delta x \rightarrow 0$) limit.  A lattice approximation to the 
left-hand side of Equation~\ref{eq-schrodinger1d} tells us how the Hamiltonian 
operator $\oper{H}$ affects the amplitudes of the state $\ket{\psi}$:
\begin{equation}
	\bra{n} \oper{H} \ket{\psi} = - \frac{1}{2\mu} \left ( \frac{\psi_{n+1} - 2 \psi_{n} + \psi_{n-1}}{\Delta x^2}\right )
		+ U(x_{n}) \psi_n .
\end{equation}
Therefore, the Hamiltonian matrix elements are
\begin{equation}
	H_{nm} = - \frac{1}{2 \mu} \left ( \frac{ \delta_{n+1,m} - 2 \delta_{nm} + \delta_{n-1,m} }{\Delta x^2} \right )
		+ U(x_{n}) \delta_{nm} .
	\label{eq-hamiltonian1dmatrix}
\end{equation}
For the pure state $\ket{\psi}$ the matrix elements of the density operator are 
$\rho_{nm} = \psi_{n} \conj{\psi_{m}}$.

From Equation~\ref{eq-hamiltonian1dmatrix}, we can see that the only non-zero 
probability currents $J_{nm}$ are those for which $n$ and $m$ are adjacent
lattice sites.  The net rightward probability current between sites $n$ and $n+1$ 
is
\begin{eqnarray}
	J_{n+1 , n} 
	& = & \frac{1}{i} \left ( H_{n+1 , n} \psi_{n} \conj{\psi_{n+1}} -
		\psi_{n+1} \conj{\psi_{n}} H_{n , n+1} \right ) \nonumber \\
	& = & - \frac{1}{2 \mu i \Delta x^2} \left (\psi_{n} \conj{\psi_{n+1}} - \psi_{n+1} \conj{\psi_{n}} \right ) 
		\nonumber \\
	& = & - \frac{1}{2 \mu i \Delta x} \bigg[ \Psi (x_{n}) \conj{\Psi}(x_{n+1}) 
	\nonumber \\ & & \quad \quad \quad \quad 
	- \Psi (x_{n+1}) \conj{\Psi(x_{n})} \bigg].
\end{eqnarray}
Given the lattice approximation for the derivative at $x_{n}$
\begin{equation}
	\dispderiv{\Psi}{x} = \frac{\Psi(x_{n+1}) - \Psi(x_{n})}{\Delta x} ,
\end{equation}
we obtain
\begin{eqnarray}
	J_{n+1,n} 
	& = & - \frac{1}{2 \mu i} \Bigg[ \Psi(x_{n})  \left ( \conj{\Psi}(x_{n}) + \dispderiv{\conj{\Psi}}{x} \Delta x \right ) 
		\nonumber \\ & & \qquad \qquad
		-  \conj{\Psi} (x_{n})  \left ( \Psi(x_{n}) + \dispderiv{\Psi}{x} \Delta x \right ) \Bigg]
		\nonumber \\ 
	& = & \frac{1}{2 \mu i} \left ( \conj{\Psi} \dispderiv{\Psi}{x} - \Psi \dispderiv{\conj{\Psi}}{x} \right ) ,
\end{eqnarray}
where the wave functions and derivatives are evaluated at $x_{n}$.  
This is exactly the lattice approximation for $J(x_{n})$ as defined in Equation~\ref{eq-currentdensity1d}
above.

The probability density $\mathcal{P} (x_{n}) = \absolute{\Psi(x_{n})}^{2} = P_{n} / \Delta x$, 
so its time derivative is
\begin{eqnarray}
	\dispderiv{\mathcal{P}}{t} = \frac{\dot{P}_{n}}{\Delta x}
	& = & \frac{J_{n,n+1} + J_{n-1,n}}{\Delta x} \nonumber \\ 
	&  = &  - \frac{J(x_{n}) - J(x_{n-1})}{\Delta x},
\end{eqnarray}
which yields Equation~\ref{eq-continuity} as $\Delta x \rightarrow 0$.  
Thus, our discrete probability current $J_{nm}$ between sites of the lattice 
reduces to the conventional probability current density in the continuous limit.

\section{Lindblad systems}

An open quantum system can exchange energy and information with its environment,
leading to correlations between system and environment.
However, it may happen that these correlations
are rapidly ``hidden'' in distant or inaccessible parts of the environment, so that on
intermediate timescales the system and environment remain effectively uncorrelated.
Under these circumstances, the time evolution of the quantum system is approximately
Markovian and may be described by a quantum master equation 
known as the Lindblad equation \cite{SchumacherWestmoreland2010}:
\begin{equation}
	\dot{\doper{\rho}} = \frac{1}{i \hbar} \left [  \oper{H}, \doper{\rho} \right ] 
		+ \sum_{\alpha} \left ( \oper{L}_{\alpha} \doper{\rho} \oper{L}_{\alpha}^{\dagger}
		- \onehalf \{ \oper{L}_{\alpha}^{\dagger} \oper{L}_{\alpha}, \doper{\rho} \} \right ),
\label{eq-lindblad}
\end{equation}
where $\{ \cdots , \cdots \}$ is the anticommutator.
The $\oper{L}_{\alpha}$ operators represent interactions with the environment
that lead to non-unitary evolution for $\doper{\rho}$.  (The Hamiltonian $\oper{H}$
may also include effects from these interations.)  It is sometimes convenient to
write Equation~\ref{eq-lindblad} in a more compact form
\begin{equation}
	\dot{\doper{\rho}} = \frac{1}{i \hbar} \left [  \oper{H}, \doper{\rho} \right ] 
	+ \mathcal{L}(\doper{\rho}) ,
	\label{eq-lindbladcompact}
\end{equation}
where $\mathcal{L}$ is the map defined by the Lindblad operators $\oper{L}_{\alpha}$.

The additional terms in Equation~\ref{eq-lindblad} require us to generalize the
definition of the probability current $J_{nm}$.  The simplest approach is as 
follows.  We first rewrite $\mathcal{L}$ in a more symmetrical form:
\begin{equation}
	\mathcal{L}(\doper{\rho})
	= \onehalf  \sum_{\alpha} \left ( \oper{L}_{\alpha} \doper{\rho} \oper{L}_{\alpha}^{\dagger}
		+ \oper{L}_{\alpha} \doper{\rho} \oper{L}_{\alpha}^{\dagger}
		- \oper{L}_{\alpha}^{\dagger} \oper{L}_{\alpha} \doper{\rho} 
		- \doper{\rho} \oper{L}_{\alpha}^{\dagger} \oper{L}_{\alpha} \right ).
\end{equation}
We now fix a basis and calculate $\dot{P}_{n} = \bra{n} \dot{\doper{\rho}} \ket{n}$.
The Hamiltonian term produces a ``Hamiltonian current'' exactly as before.  
Additional terms due to the $\mathcal{L}(\doper{\rho})$ map can be found by introducing a complete
basis $\{ \ket{m} \}$ at appropriate points within the operator products.  This procedure yields
\begin{equation}
	J_{nm} = J_{nm}^{(H)} + \bigsum{\alpha} J_{nm}^{(\alpha)}
	\label{eq-jdeflindblad}
\end{equation}
where $J_{nm}^{(H)}$ is the Hamiltonian current defined previously in Equation~\ref{eq-jdefhamiltonian}
and
\begin{eqnarray}
	J_{nm}^{(\alpha)} 
	& = & \onehalf \Big( 
		\left ( \oper{L}_{\alpha} \doper{\rho} \right )_{nm} \left ( \oper{L}_{\alpha}^{\dagger} \right )_{mn}
		\nonumber \\ & & \quad 
	   	+ \left (\oper{L}_{\alpha} \right )_{nm} \left (\doper{\rho} \oper{L}_{\alpha}^{\dagger} \right )_{mn} 
		\nonumber \\ & & \quad 
		- \left (\oper{L}_{\alpha}^{\dagger} \right )_{nm} \left ( \oper{L}_{\alpha} \doper{\rho} \right )_{mn}
		\nonumber \\ & & \quad 
		- \left ( \doper{\rho} \oper{L}_{\alpha}^{\dagger} \right )_{nm} \left (\oper{L}_{\alpha}\right )_{mn}
		\Big)
	\label{eq-jalphadef}
\end{eqnarray}
represents the contribution to the probability current produced by a single Lindblad operator 
$\oper{L}_{\alpha}$.

We can readily verify that $J_{nm}$ is real and antisymmetric, and that the Lindblad equation 
guarantees that Equation~\ref{eq-probabilityflow} holds.

Note that our probability current is a linear functional of the state $\doper{\rho}$.
This is a reasonable property for $J_{nm}$.  
Suppose the density operator $\doper{\rho}$ is a probabilistic
mixture of two states:  $\doper{\rho} = p_{1} \doper{\rho}_{1} + p_{2} \doper{\rho}_{2}$.  Then
all probabilities are also mixtures:
\begin{equation}
	P_{n} = p_{1} \bra{n} \doper{\rho}_{1} \ket{n} + p_{2} \bra{n} \doper{\rho}_{2} \ket{n} .
\end{equation}
It thus makes sense that the probability currents $J_{nm}$, which are related to
the rates of change $\dot{P}_{n}$, behave in the same way under mixtures:
\begin{equation}
	J_{nm} = p_{1} J_{nm,1} + p_{2} J_{nm,2} ,
\end{equation}
as Equation~\ref{eq-jdeflindblad} implies.

The representation of open system evolution in Equation~\ref{eq-lindblad} is not
unique.  Suppose we replace the $\oper{L}_{\alpha}$ 
operators with modified $\oper{M}_{\beta}$ operators, where
\begin{equation}
	\oper{M}_{\beta} = \sum_{\alpha} U_{\beta \alpha} \oper{L}_{\alpha},
\end{equation}
and the $U_{\beta \alpha}$ coefficients are the elements of a unitary matrix.
The resulting Lindblad equation is exactly equivalent to the original.  To see
this, we first note that
\begin{equation}
	\sum_{\beta} \conj{U_{\beta \alpha}} U_{\beta \gamma}
	= \sum_{\beta} \left ( U^{\dagger} \right )_{\alpha \beta} U_{\beta \gamma} = \delta_{\alpha \gamma}.
\end{equation}
Using this it is not hard to verify that 
\begin{eqnarray}
	\sum_{\beta} \oper{M}_{\beta}^{\dagger} \oper{M}_{\beta} & = &
	\sum_{\alpha} \oper{L}_{\alpha}^{\dagger} \oper{L}_{\alpha}
	\nonumber \\
	\sum_{\beta} \oper{M}_{\beta}^{\dagger} \oper{\rho} \oper{M}_{\beta} & = &
	\sum_{\alpha} \oper{L}_{\alpha}^{\dagger} \oper{\rho} \oper{L}_{\alpha}
\end{eqnarray}
Therefore the open system dynamics described by Equation~\ref{eq-lindblad}
will be unchanged by the replacement of the $\oper{L}$-operators with 
$\oper{M}$-operators.  The two
sets of operators, $\{ \oper{L}_{\alpha} \}$ and $\{ \oper{M}_{\beta} \}$,
are sometimes called two {\em unravellings} of the same Lindblad master 
equation.  

The definition of $J_{nm}$ in Equation~\ref{eq-jdeflindblad} has the same
general structure as the Lindblad equation itself.  Each term involving $\oper{L}_{\alpha}$
also involves $\oper{L}_{\alpha}^{\dagger}$, and the index $\alpha$ is summed over.
Therefore, the value of the probability current $J_{nm}$ is independent of our choice of 
unravelling of the master equation.  The individual $J_{nm}^{(\alpha)}$ terms, of 
course, may be different for different unravelings, but their sum will be the same.

\section{A flow paradox}

Consider a system governed by a classical master equation, Equation~\ref{eq-classicalmastereqn},
with classical probability current $J_{nm} = W_{nm} P_{m} - W_{mn} P_{n}$.  
If both $P_{n} = 0$ and $P_{m} = 0$, the current $J_{nm}$ from $m$ to $n$ must be zero as well.
If only $P_{n} = 0$, we can still have a nonzero $J_{nm}$; however, since the conditional
rates $W_{nm} \geq 0$ for all pairs of states, we can conclude that $J_{nm} \geq 0$ in this case.
If a state $n$ has zero probability, the net flow of probability between $n$ and some other state
$m$ can only be directed toward $n$.

By contrast, for a quantum system governed by the Schr\"{o}dinger equation 
(Equation~\ref{eq-schrodingerrho}),
the quantum probability flow $J_{nm} = 0$ if {\em either} $P_{n} = 0$ or $P_{m} = 0$.

What about the generalized definition of $J_{nm}$ for Lindblad evolution (Equation~\ref{eq-jdeflindblad})?
Consider a qubit system $Q$ 
with basis states $\ket{0}$ and $\ket{1}$, whose dynamics may be described by a
single Lindblad operator $\oper{L} = \sqrt{\lambda} \outerprod{0}{1}$, where $\lambda$ is a constant.  
(This might describe a decay process in a two level atom 
from the excited state $\ket{1}$ to the ground state $\ket{0}$.)
If at a particular moment the state of $Q$ is described by the density
operator $\doper{\rho} = \proj{1}$, the probability current $J_{01} = \lambda$
(and so $\dot{P}_{0} = \lambda$), even though $P_{0} = 0$.

This fact presents us with a puzzle.  We may consider a composite system 
comprising $Q$ and its environment $E$.  The complete system $QE$ 
is informationally isolated, so that the evolution 
of its joint state is described by the Schr\"{o}dinger equation.
If we choose basis states $\{ \ket{n} \}$ for $Q$ and $\{ \ket{a} \}$ for
$E$, the joint probability is $P_{na}$, and the total
probability for $Q$-state $n$ is 
\begin{equation}
	P_{n} = \sum_{a} P_{na}.
\end{equation}
Thus, if $P_{n} = 0$, then it must be that all of the probabilities $P_{na} = 0$.  
From this it follows that all probability currents $J_{na,mb} = 0$ 
in this composite system.

What is the relation between $J_{na,mb}$ for $QE$ and $J_{nm}$ for $Q$ alone? 
This can be seen by considering how $P_{n}$ is changing over time:
\begin{eqnarray}
	\dot{P}_{n} = \sum_{a} \dot{P}_{na} & = & \sum_{a} \left ( \sum_{mb} J_{na,mb} \right )
	\nonumber \\
	& = & \sum_{m} \left ( \sum_{ab} J_{na,mb} \right ) .
\end{eqnarray}
Hence we identify
\begin{equation}
	J_{nm} = \sum_{ab} J_{na,mb} .
\end{equation}
We may conclude that, if $P_{n} = 0$ for the subsystem $Q$,
the rate of change $\dot{P}_{n} = 0$ too.

However, as our decaying qubit example shows, this is not necessarily true
for evolution described by the Lindblad equation.  We may have $P_{n} = 0$
but $J_{nm}$ and $\dot{P}_{n}$ both nonzero.  How can this behavior arise
if the global state is evolving unitarily?

The answer comes from recognizing that the Lindblad equation can only be an
{\em approximation} to the exact evolution of a subsystem state.  Consider
another system, an isolated qubit with internal Hamiltonian 
\begin{equation}
	\oper{H} = - \omega \oper{Y} = i \omega \left ( \outerprod{0}{1} - \outerprod{1}{0} \right ) .
\end{equation}
The state of the system happens to be the pure state $\ket{\psi} = \epsilon \ket{0}
+ \sqrt{1-\epsilon^2} \ket{1}$, where $\epsilon \ll 1$.  Then
\begin{eqnarray}
	\doper{\rho} = \proj{\psi} & = & \epsilon^2 \proj{0} + \epsilon \sqrt{1-\epsilon^2} \outerprod{0}{1} \nonumber \\
		& & + \epsilon \sqrt{1-\epsilon^2} \outerprod{1}{0} + (1 - \epsilon^2) \proj{1}{1} .
\end{eqnarray}
The probability $P_{0} = \rho_{00} = \epsilon^2$.  The probability current is
\begin{equation}
	J_{01} = \frac{1}{i} (H_{01} \rho_{10} -  \rho_{01}H_{10}) = 2 \omega \epsilon \sqrt{1-\epsilon^2} .
\end{equation}
To lowest order in $\epsilon$, $J_{01} \approx 2 \omega \epsilon$.  Thus, if we 
consider an approximation in which we keep $\oforder{\epsilon}$ terms 
but discard $\oforder{\epsilon^2}$ terms, we
will say that $P_{0} = 0$ but $J_{01} \neq 0$ in our approximation.

Suppose therefore that the environment $E$ has a Hilbert space of large dimension 
$D$.  (Such a complex environment would be required in order to continually ``hide''
its correlations with $Q$.)  
For a given system state $Q$, we can consider a state for which the $\ket{n,a}$
basis states have small (that is, $\oforder{\epsilon}$) amplitudes.
Then $P_{na} \sim \epsilon^2$ for each $a$, and thus
\begin{equation}
	P_{n} = \sum_{a} P_{na} \sim D \epsilon^2 .
\end{equation}
On the other hand, we may expect $J_{na,mb} \sim \epsilon$, and so
\begin{equation}
	J_{nm} = \sum_{ab} J_{na,mb} \sim D^2 \epsilon .
\end{equation}
If $\epsilon$ is small and $D$ is large, it may be that $D \epsilon^2$ is
negligible, but $D^2 \epsilon$ is finite.  This is why the approximate
Lindblad evolution can have $P_{n} = 0$ but $J_{nm} \neq 0$.

In this way, Lindblad evolution behaves more like a classical master equation.
On the other hand, the quantum case can exhibit some distinctly non-classical
features.  Consider a system with $\dim \hilbert = 3$ and basis
states $\ket{1}$, $\ket{2}$ and $\ket{3}$.  The internal system
Hamiltonian is zero, and the non-unitary part of the evolution is
given by a single Lindblad operator
\begin{equation}
	\oper{L} = \sqrt{\lambda} \left ( \outerprod{3}{1} - 2 \outerprod{3}{2} \right ).
\end{equation}
(That is, the only non-zero matrix elements are $L_{31} = 1$ and $L_{32} = -2$.)
Our state $\doper{\rho} = \proj{\psi}$, 
where $\ket{\psi} = \frac{1}{\sqrt{2}} \left ( \ket{1} + \ket{2} \right )$.
From this we can calculate that $P_{3} = \rho_{33} = 0$ and
\begin{equation}
	J_{31} = -\frac{\lambda}{4} \quad \mbox{and} \quad J_{32} = + \frac{\lambda}{2} .
\end{equation}
Not only are these non-zero, but $J_{13} = -J_{31}$ (the net probability 
current from state 3 to state 1) is actually positive.  We may have a positive probability
current from a state of zero probability!  (No actual paradox is involved, since
$J_{31} + J_{32} > 0$, so the net probability current is {\em into} the
zero-probability state 3.)

In spite of this surprising situation, it remains true that
if both $P_{n}$ and $P_{m}$ are zero, the current $J_{nm} = 0$ also,
even for Lindblad evolution.  The simplest way to prove this is to note that
if $P_{n} = \rho_{nn} = 0$, then $\rho_{nk} = \conj{\rho_{kn}} = 0$ for all $k$,
and thus for any $\oper{L}$
\begin{equation}
	\left ( \doper{\rho} \oper{L} \right )_{nm} = \sum_{k} \rho_{nk} L_{km} = 0
	\quad \mbox{and} \quad
	\left (\oper{L}  \doper{\rho} \right )_{mn} = \sum_{k} L_{mk} \rho_{kn} = 0 .
\end{equation}
Each term in Equation~\ref{eq-jalphadef} involves a factor of this type, and so
$J_{nm}^{(\alpha)} = 0$ for every Lindblad operator $\oper{L}_{\alpha}$.  (The
Hamiltonian part of the current $J_{nm}^{H} = 0$ as well, by the arguments
given above.)

\section{Heat and and probability flow}

The Hamiltonian $\oper{H}$ is the energy operator for a system, and the 
system's mean energy is the expectation
\begin{equation}
	\ave{E} = \tr \doper{\rho} \oper{H} .
\end{equation}
The value of $\ave{E}$ can change in two distinct ways.  The system state
$\rho$ may change, or the Hamiltonian $\oper{H}$ may change due to the
modification of one or more external parameters (e.g., a change in an 
externally applied magnetic field).  Thus, we may write
\begin{equation}
	\frac{d}{dt} \ave{E} = \tr \dot{\doper{\rho}} \oper{H} + \tr \doper{\rho} \dot{\oper{H}} . 
\end{equation}
These two terms correspond to thermodynamic {\em heat} and {\em work}, 
respectively \cite{SchumacherWestmoreland2010}.  That is, if we let $\mathcal{P}_{Q}$ and $\mathcal{P}_{W}$
denote the rates at which heat and work are transferred to the system, we
may write
\begin{equation}
	\mathcal{P}_{Q} = \tr \dot{\doper{\rho}} \oper{H} 
	\quad \mbox{and} \quad
	\mathcal{P}_{W} = \tr \doper{\rho} \dot{\oper{H}} .
\end{equation}

Consider a system interacting with its environment so that its evolution
is described by the Lindblad equation (Equation~\ref{eq-lindbladcompact}).
Note that the Hamiltonian $\oper{H}$ and the Lindblad operators $\oper{L}_{\alpha}$
may depend on time.  We find that
\begin{eqnarray}
	\mathcal{P}_{Q}
	& = & \tr \dot{\doper{\rho}} \oper{H}  \nonumber \\
	& = & \tr \left ( \frac{1}{i} [\oper{H},\doper{\rho}] \right ) \oper{H} + \tr \mathcal{L}(\doper{\rho}) \oper{H} 
		\nonumber \\
	& = & \frac{1}{i} \tr \left ( \oper{H} \doper{\rho} \oper{H} - \doper{\rho} \oper{H}^2 \right )
		+ \tr \mathcal{L}(\doper{\rho}) \oper{H} \nonumber \\
	& = & \tr \mathcal{L}(\doper{\rho}) \oper{H} ,
\end{eqnarray}
since $\tr \oper{H} \doper{\rho} \oper{H} = \tr \doper{\rho} \oper{H} \oper{H}$.  Only the 
non-unitary part of the dynamics described by the map $\mathcal{L}$ affects the
heat rate $\mathcal{P}_{Q}$.  This map is absent if the system is informationally isolated,
so $\mathcal{P}_{Q} = 0$.
Thus, energy changes in such a system can only involve work 
done on the system, not heat transfer to the system.

In this section and the next, we will explore the relation of 
$\mathcal{P}_{Q}$ and $\mathcal{P}_{W}$ to the probability current $J_{nm}$.
For our basis we choose the energy eigenbasis, in which the
eigenstate $\ket{n}$ has energy $E_{n}$ and the Hamiltonian
operator is diagonal:
\begin{equation}
	\oper{H} = \sum_{n} E_{n} \proj{n} .
\end{equation}
This has the advantage that the Hamiltonian part of the current
$J_{nm}^{(H)} = 0$.  It has the disadvantage that the Hamiltonian
may vary with time.  Thus, $P_{n}$ and $J_{nm}$ will be defined 
with respect to a basis $\{ \ket{n} \}$ that is not fixed, but itself
has some time-dependence.  As we will see below, this requires us
to introduce additional terms in some of our expressions.

Heat transfer, however, can be analyzed with the expressions
already in hand, simply by adopting the instantaneous energy
eigenbasis $\{ \ket{n} \}$ and defining $J_{nm}$ according to
Equation~\ref{eq-jdeflindblad}.
We obtain
\begin{eqnarray}
	\mathcal{P}_{Q}
	& = & \sum_{n} E_{n} \bra{n} \mathcal{L}(\doper{\rho}) \ket{n} \nonumber \\
	& = & \sum_{n} \sum_{\alpha} E_{n} \Big(
		(\oper{L}_{\alpha} \doper{\rho} \oper{L}_{\alpha}^{\dagger})_{nn}
		\nonumber \\ & & \qquad \qquad
		- \onehalf (\oper{L}_{\alpha}^{\dagger} \oper{L}_{\alpha} \doper{\rho})_{nn}
		- \onehalf (\doper{\rho} \oper{L}_{\alpha}^{\dagger} \oper{L}_{\alpha})_{nn}
		\Big) \nonumber \\
	& = & \sum_{nm} \sum_{\alpha} E_{n} J_{nm}^{(\alpha)} 
\end{eqnarray}
where the $J_{nm}^{(\alpha)}$ currents are defined in Equation~\ref{eq-jalphadef}.
Since the Hamiltonian current $J_{nm}^{(H)} = 0$, we may simply write
\begin{equation}
	\mathcal{P}_{Q} =  \sum_{nm} E_{n} J_{nm} .
	\label{eq-heatrateasymmetric}
\end{equation}
By reindexing the last sum, we note that
\begin{equation}
	\sum_{nm} E_{n} J_{nm} = \sum_{nm} E_{m} J_{mn} = - \sum_{nm} E_{m} J_{nm}  .
\end{equation}
Therefore we may write
\begin{equation}
	\mathcal{P}_{Q} = \onehalf \sum_{nm} \left ( E_{n} - E_{m} \right ) J_{nm}  .
	\label{eq-heatrate}
\end{equation}

If the eigenbasis $\{ \ket{n} \}$ of the Hamiltonian is time-independent,
the interpretation of Equation~\ref{eq-heatrate} is straightforward.
A transition of the system from state $m$ to state $n$ produces a
change in system energy $\Delta E = E_{n} - E_{m}$.  The current
$J_{nm}$ is, in a probabilistic sense, the net rate at which such 
transitions are happening.  The factor of $1/2$ arises because the
double sum counts each distinct pair of states twice.  We could in fact
rewrite Equation~\ref{eq-heatrate} as
\begin{equation}
	\mathcal{P}_{Q} = \sum_{n>m} \left ( E_{n} - E_{m} \right ) J_{nm} .
\end{equation}

We can also write Equation~\ref{eq-heatrate} as
\begin{equation}
	\mathcal{P}_{Q} = \sum_{\alpha} \left ( \onehalf \sum_{nm} \left ( E_{n} - E_{m} \right ) 
	  J_{nm}^{(\alpha)} \right )  = \sum_{\alpha} \mathcal{P}_{Q}^{(\alpha)}.
	  \label{eq-heatratebyalpha}
\end{equation}
The different Lindblad operators $\oper{L}_{\alpha}$ may represent different
physical processes.  For example, it may be that our system of interest is
interacting with two different parts of the environment---heat baths at 
different temperatures, perhaps.  We may be able to partition the set of Lindblad
operators into two groups, one for each external system.  By dividing the terms of
the sum in Equation~\ref{eq-heatratebyalpha}, we can determine
separate heat transfer rates from the two external systems.

\section{Work and the rotating eigenbasis}

Now we turn to the situation in which the Hamiltonian $\oper{H}$, 
which is both the instantaneous energy observable and the generator 
of the unitary part of the time evolution in Equation~\ref{eq-lindblad},
can itself vary over time.  We write\footnote{We are assuming that both
the eigenvalues and eigenstates of $\oper{H}$ vary in a smooth and
continuous way.  This is not necessarily true for the eigenstates.
Consider a qubit with $\oper{H}(t) = at^2 \oper{Z}$ for $t \leq 0$ and
$\oper{H} = at^2 \oper{X}$ for $t \geq 0$.  At $t=0$, $\dot{\oper{H}}$ is
well-defined (and equals zero), but the eigenbasis 
for $\oper{H}$ changes instantaneously from the
$\oper{Z}$-basis to the $\oper{X}$ basis.  This can happen because the
energy eigenvalues are degenerate at $t=0$.  Any small perturbation to
$\oper{H}$ that lifts this degeneracy will also remove the discontinuity.
Since the discontinuous case relies on a special exact degeneracy, 
we adopt the assumptions of Equation~\ref{eq-hdotsimple} as
physically reasonable in the generic case.}
\begin{eqnarray}
	\dot{\oper{H}} 
	& = & \frac{d}{dt} \left ( \sum_{n} E_{n} \proj{n} \right ) \nonumber \\
	& = & \sum_{n} \dot{E}_{n} \proj{n} + \sum_{n} E_{n} \left ( \frac{d}{dt} \proj{n} \right ) .
	\label{eq-hdotsimple}
\end{eqnarray}
The first sum yields an operator $\oper{R}$:
\begin{equation}
	\oper{R} = \sum_{n} \dot{E}_{n} \proj{n} .
\end{equation}
This operator, which commutes with $\oper{H}$, describes how the 
energy eigenvalues are changing.

How do the basis states $\ket{n}$ change over time?  Since the basis
remains orthonormal, the basis vectors must change unitarily.  This 
unitary evolution is generated by a Hermitian operator $\oper{K}$, 
which appears in a ``Schr\"{o}dinger'' equation for $\ket{n(t)}$:
\begin{equation}
	\oper{K} \ket{n(t)} = i \frac{d}{dt} \ket{n(t)} ,
\end{equation}
which can also be written as
\begin{equation}
	\frac{d}{dt} \proj{n} = \frac{1}{i} [ \oper{K}, \proj{n} ] .
\end{equation}
This lets us simplify the second sum in Equation~\ref{eq-hdotsimple}:
\begin{eqnarray}
	\sum_{n} E_{n} \left ( \frac{d}{dt} \proj{n} \right )
	& = & \frac{1}{i} \sum_{n} E_{n} \, [ \oper{K}, \proj{n} ] 
	\nonumber \\ & = & \frac{1}{i} [\oper{K}, \oper{H}] .
\end{eqnarray}
Therefore we can write
\begin{equation}
	\dot{\oper{H}} = \oper{R} + \frac{1}{i} [ \oper{K} , \oper{H} ],
	\label{eq-hdot}
\end{equation}
where the operators $\oper{R}$ and $\oper{K}$ describe the changes in
the eigenvalues and eigenstates of $\oper{H}$, respectively.

How does the change in basis---a change in our ``reference frame'' in the Hilbert 
space---affect probability currents?  At any
moment, the probability of the $n$th state is $P_n = \bra{n} \doper{\rho} \ket{n}$,
as before; but now both $\doper{\rho}$ and $\ket{n}$ may be changing:
\begin{equation}
	\dot{P}_{n} = \bra{n} \dot{\doper{\rho}} \ket{n} + \frac{1}{i} \tr \rho \, [\oper{K},\proj{n}] .
\end{equation}
By cyclically permuting terms in the trace, we can transform this to
\begin{equation}
	\dot{P}_{n} = \bra{n} \left ( \dot{\doper{\rho}} - \frac{1}{i} [\oper{K},\doper{\rho}] \right ) \ket{n} .
\end{equation}
As far as the changes in $P_{n}$ are concerned, the effect of the change in
basis is the same as introducing a new term $-\oper{K}$ into the Hamiltonian.
Thus, we introduce a {\em frame current}
\begin{equation}
	I_{nm} = - \frac{1}{i} \left (K_{nm} \rho_{mn} - \rho_{nm} K_{mn} \right ) ,
\end{equation}
and write
\begin{equation}
	\dot{P}_{n} = \sum_{m} \left ( J_{nm} + I_{nm} \right ) .
	\label{eq-pdotwithframe}
\end{equation}
The probability currents $J_{nm}$ are related to changes in the quantum state $\doper{\rho}$
and are defined as in Equations~\ref{eq-jdefhamiltonian} and
\ref{eq-jdeflindblad}.  To account for all changes in probability, including those
produced by our shifting basis, we need to include
the frame current $I_{nm}$ as well.

The rate at which work is done on the system is
\begin{eqnarray}
	\mathcal{P}_{W} & = & \tr \doper{\rho} \dot{\oper{H}} \nonumber \\
	& = & \tr \doper{\rho} \oper{R} + \frac{1}{i} \tr \doper{\rho} [\oper{K}, \oper{H}] \nonumber \\
	& = & \tr \doper{\rho} \oper{R} - \frac{1}{i} \tr [\oper{K},\doper{\rho}] \oper{H} .
\end{eqnarray}
In the energy eigenbasis this becomes
\begin{equation}
	\mathcal{P}_{W} = \sum_{n} P_{n} \dot{E}_{n} + \sum_{nm} E_{n} I_{nm} .
\end{equation}
Using the antisymmetry of the frame current $I_{nm}$, we obtain
\begin{equation}
	\mathcal{P}_{W} = \sum_{n} P_{n} \dot{E}_{n} + \frac{1}{2} \sum_{nm} (E_{n} - E_{m}) I_{nm} .
	\label{eq-workrate}
\end{equation}

The second term in Equation~\ref{eq-workrate} resembles the heat transfer rate in
Equation~\ref{eq-heatrate}, but it involves the frame current $I_{nm}$ rather than
$J_{nm}$.  It tells us that work can be done on a quantum system either by shifting
the energy levels (the $\dot{E}_{n}$ term) or by rotating the energy basis.  Consider
a spin in an external magnetic field.  We may do work on the spin either by changing
the magnitude of the field (which changes the energy eigenvalues) or by rotating
the field to a new spatial direciton (which changes the energy eigenstates).

To summarize, we have found that the mean energy of the system changes by
\begin{eqnarray}
	\frac{d}{dt} \ave{E} 
	& = & \overbrace{\left ( \onehalf \sum_{nm} \left ( E_{n} - E_{m} \right ) J_{nm} \right )}^{\mathcal{P}_{Q}}
	\nonumber \\ &  & 
		+ \underbrace{\left ( \sum_{n} P_{n} \dot{E}_{n} + \frac{1}{2} \sum_{nm} (E_{n} - E_{m}) I_{nm} \right )}_{\mathcal{P}_{W}} 
\end{eqnarray}
where $\mathcal{P}_{Q}$ and $\mathcal{P}_{W}$ are the rates of heat and work
energy transfers to the system.

\section{Remarks}

We have proposed particular definitions for the probability current
$J_{nm}$.  For informationally isolated systems evolving according
to the Schr\"{o}dinger equation, the definition in
Equation~\ref{eq-jdefhamiltonian} suffices; for systems evolving
via a Lindblad master equation, we have the more general 
definition in Equation~\ref{eq-jdeflindblad}.  These definitions
share a number of general properties:
\begin{itemize}
	\item  $J_{nm}$ is real.
	\item  $J_{nm}$ is antisymmetric in $n$ and $m$.
	\item  Given a fixed basis set $\{ \ket{n} \}$, $\dot{P}_{n} = \bigsum{m} J_{nm}$.
	\item  $J_{nm}$ is independent of the particular unraveling of the
		quantum master equation via Lindblad operators.
	\item  $J_{nm}$ is a linear functional of the state $\doper{\rho}$.
\end{itemize}
We must note, however, that these properties are shared by many
other definitions.  Suppose $a_{nm}$ is an arbitrary real 
antisymmetric matrix of coefficients such that every row sums to
zero, and let $\oper{B}$ be a Hermitian operator.  Then the revised 
expression
\begin{equation}
	J_{nm}' = J_{nm} + a_{nm} \left ( \tr \doper{\rho} \oper{B} \right )
	\label{eq-jprimedef}
\end{equation}
yields probability currents that satisfy all of the general properties 
we have listed.

A similar ambiguity exists for the probability current density $\vec{J}$
for a particle moving in three dimensions.  Together with the probability
density $\mathcal{P} = \absolute{\Psi}^{2}$, the current density
satisfies the continuity equation:
\begin{equation}
	\dispderiv{\mathcal{P}}{t} = - \vec{\nabla} \cdot \vec{J} .
\end{equation}
If we define a new current density $\vec{J}' = \vec{J} + \vec{K}$,
where $\vec{\nabla} \cdot \vec{K} = 0$, the continuity equation
for probability is still satisfied.

In fact, many of our results work equally well for the modified
probability current of Equation~\ref{eq-jprimedef}.  Consider
the heat rate relation from Equation~\ref{eq-heatrateasymmetric}.
If we replace $J_{nm}$ by $J_{nm}'$, we obtain
\begin{eqnarray}
	\sum_{nm} E_{n} J_{nm}' 
	& = & \sum_{nm} E_{n} J_{nm} + \sum_{nm} E_{n} a_{nm} \left ( \tr \doper{\rho} \oper{B} \right )
	\nonumber \\
	& = & \mathcal{P}_{Q} + \sum_{n} E_{n} ( \tr \doper{\rho} \oper{B} ) \left ( \sum_{m} a_{nm} \right )
	\nonumber \\
	& = & \mathcal{P}_{Q} .
\end{eqnarray}
The modified currents are related to heat transfer in exactly the same
way as before.

Since the definitions in Equation~\ref{eq-jdefhamiltonian}
and \ref{eq-jdeflindblad} are so closely connected to 
the dynamical equations, we believe that they are the simplest 
choices for the probability currents $J_{nm}$.
They also have some reasonable properties 
that are not shared by some 
of the alternatives given by Equation~\ref{eq-jprimedef}.  For instance,
under our definitions, if $P_{n}$ and $P_{m}$ are both zero, then
$J_{nm} = 0$; but it is easy to come up with examples for which
$J_{nm}' \neq 0$.  This raises an unresolved question:  Can
we devise a reasonable, physically motivated set of general 
properties for the probability currents $J_{nm}$ that uniquely
determine their definition?

In spite of the apparent arbitrariness of the definitions given here,
we anticipate that the probability currents $J_{nm}$ will be a 
useful tool for understanding the dynamics of quantum systems
with finite-dimensional Hilbert spaces.  The connections between
probability currents and thermodynamic work and heat also suggest
that they will be helpful in analyzing the function of quantum
thermodynamic systems like the small thermal machines analyzed 
in \cite{LindenPopescuSkrzypczyk2010}.

The authors wish to express their gratitude for many useful conversations 
on these and related problems with Lidia del Rio, Michael Morgan, 
Sandu Popescu, and Tony Short.
AN was supported by the Kenyon Summer Science Scholar
program in the summer of 2013.  
BWS and MDW acknowledge the support of the Foundational 
Questions Institute (FQXi), via grant FQXi-RFP-1517.

\bibliography{quantum}
\bibliographystyle{apsrev4-1}

\end{document}